# Controllable Anomalous Dispersion and Group Index Nulling via Bi-Frequency Raman Gain in Rb Vapor for application to Ultraprecision Rotation Sensing

**G.S. Pati, Renu Tripathi, M. Messall, K. Salit and M.S. Shahriar**

*Department of Electrical Engineering and ComputerScience,*

*Northwestern University, Evanston IL 60208*

We have recently proposed [9], the use of 'fast-light' media to obtain ultrahigh precision rotation sensing capabilities. The scheme relies on producing a critically anomalous dispersion (CAD), in a suitable dispersive medium, which is introduced in the arms of a Sagnac interferometer. We present here an experimental investigation of the anomalous dispersion properties of bi-frequency Raman gain in Rb vapor, with the goal of using this medium for producing the CAD condition. A heterodyne phase measurement technique is used to measure accurately the index variation associated with the dispersion. The slope of the negative linear dispersion (or group index) is experimentally varied by more than two orders of magnitude while changing the frequency separation between pump fields, responsible for producing gain. Using this result, we have identified the experimental parameters for achieving a null value of the group index, corresponding to the CAD condition necessary for enhanced rotational sensitivity.

Resonant dispersion phenomena that originate due to optically induced coherence between atomic levels of alkali atom vapors or rare-earth solids have caused a resurgence of interest in problems that involve light propagation in dispersive media [1-4]. Earlier experiments have reported significantly large values of the linear dispersion $\partial n/\partial \omega$ [O($10^{-8}$)] in order to demonstrate "ultra-slow light", which is promising for many applications including photon switching and quantum memory [5-7]. The motivation for our study stems from recent proposals [8,9] that have addressed the subject of light propagation in resonantly dispersive *moving* media, such as an optical gyroscope, and have predicted conditions under which extreme dispersion can play an important role in sensitivity enhancement. It has been found that the slow-light-related dispersion dramatically modifies the light-drag coefficient to enhance rotational sensitivity of a Sagnac interferometer, but only when measuring the *relative* motion. In contrast, anomalous or negative dispersion (with appropriate bounds) can enhance the sensitivity of an *absolute* rotation sensor by many orders of magnitude [9]. To realize the effect of negative dispersion in a resonator gyroscope, we are considering an atomic medium that can exhibit resonant anomalous dispersion with $|\partial n/\partial \omega| \approx (n_o/\omega_o)$ corresponding to a group index $0 < n_g << 1$, where $n_o$ is the mean value of medium index and $\omega_o$ is the center frequency for resonant dispersion. A higher sensitivity is achieved for value of $n_g$ close to its null value [9], a condition known as critically anomalous dispersion (CAD).

The goal of this paper is to report experimental identification of conditions necessary to realize a CAD medium, by using bi-frequency Raman gain in Rb vapor. In earlier experiments [4, 18], the transparent spectral gain region and the negative dispersion associated with gain doublets were used to demonstrate the superluminal propagation of optical pulses, a phenomenon known as “fast light." Anomalous dispersion gives rise to a rephasing of the different frequency components of the pulse, to move the pulse forward in time, an effect that does not violate causality. On the other hand, our application simply relies on a CAD type dispersion characteristic experienced by continuous-wave bi-directional probes in a resonator when the probe frequencies are shifted in opposite directions due to rotation [9]. Under this condition, the steep index gradient leads to a further increase in the frequency splitting between the cavity modes, and thereby, the sensitivity. Another important remark about the CAD medium is the following: most of the experiments done to date intend to achieve large value of negative $n_g$ in order to demonstrate significant pulse advancement or superluminal propagation. The typical value of linear dispersion $\partial n/\partial \omega$ reported in these experiments ranges anywhere between $10^{-11}$ to $10^{-13}$ $rad^{-1}$ sec. In contrast, a CAD medium for our application only requires a value in the range $10^{-15}$ $rad^{-1}$ sec, which is two orders of magnitude smaller. As shown here experimentally, this condition is achievable in bi-frequency Raman gain using experimentally controllable parameter.

Raman gain is observed in a three-level atomic medium under the two-photon resonance condition. It is normally achieved by applying a strong coupling laser that creates both population inversion and drives one leg of the stimulated Raman process [10-12]. Under this condition, the atomic medium behaves as a laser-like gain medium. Single-pass Raman gain with a large gain coefficient $g_o$ (~ 0.35 $cm^{-1}$) and subnatural linewidth can be achieved using a medium with modest atomic density [N ~ $O(10^{12})$] and ground-state dephasing. Experimentally, self-pumped off-resonant Raman gain as high as 12-15 dB has been observed using frequency-detuned coupling (or pump) lasers that have been tuned far below the atomic resonance in $^{85}Rb$ vapor (~ 1.2 GHz) [13]. Spectral measurements of such gain have also been performed, showing the generation of strong Raman sidebands at anti-Stokes frequencies. However, gain can also be extracted more efficiently using a relatively weak coupling laser intensity and observed over a wide range of frequency detunings, if external optical pumping is used to create a population inversion between the two ground states.

We have used the latter approach to produce bi-frequency Raman gain. It displays a steep negative dispersion in the intermediate region between the two gain regions. Our experiment shows that the dispersion slope can be varied in order to achieve a group index $n_g$ close to a zero (or null) value. This experiment has advantages when compared to the use of other absorption phenomena in coherently prepared atomic media that have recently [14, 15, 20] been used to produce negative dispersion. We have measured accurately the magnitude of linear dispersion while varying $n_g$ close to the null value using a heterodyne technique. The phase change due to the index variation associated with resonant dispersion has also been measured. The linear bandwidth and dispersion slope for the probe are estimated from these measurements. This provides information about the sensitivity, dynamic range and operational bandwidth of a CAD medium based resonator gyroscope.

Fig. 1 shows the hyperfine ground states and the excited state manifolds of the D2 line in $^{85}$Rb that constitute a Λ-type configuration for an off-resonant Raman excitation. Fig. 2 shows the schematic of our experimental setup. The probe and pump laser beams are obtained from an external cavity frequency-locked CW Ti:Sapphire laser (line width ~ 1 MHz). The frequency difference between them is matched to the ground state splitting in $^{85}$Rb (3.0357 GHz), a condition which is known as a two-photon resonance. This is achieved by frequency shifting the probe with respect to the pump using an acousto-optic modulator (AOM), which is not shown in fig.2. While observing the linewidth associated with Raman gain, the probe frequency is continuously scanned around the two-photon resonance. The beams are cross-linearly polarized.

As shown in fig.1, two pump beams of slightly different frequencies are used to generate closely spaced gain peaks, called a gain doublet, in order to produce anomalous dispersion. An incoherent beam from a tapered amplifier diode laser (fig. 2) is used for optical pumping. This beam is frequency-locked to the $5S_{1/2}$, F = 2 to $5P_{3/2}$, F′ = 3 transition in $^{85}$Rb and is orthogonally polarized with respect to the probe beam. The two pump beams and the probe are combined using two beam splitters, one polarizing, and the other non-polarizing at the input end, and made to propagate collinearly in a 10 cm long Rb vapor cell that contains a mixture of the 85 and 87 isotopes. The beams are also focused in the medium to an estimated spot size of 100 μm, which corresponds a confocal distance of approximately 5 cm. The vapor cell is magnetically shielded using two-layers of μ-metal. During the experiment, the cell is heated to nearly a steady temperature of 100°C using bifilarly wound coils that produce a negligible axial magnetic field. After passing through the vapor cell, the probe beam is separated from all the other beams by using a high extinction (50 dB) prism polarizer.

While observing the maximum probe gain, the average frequency detuning $\Delta_o$ (fig. 1) of the laser fields associated with the Λ-transition is varied by tuning the laser frequency away from the $5S_{1/2}$, F = 3 to $5P_{3/2}$ F′ = 4 transition. Maximum gain is obtained when $\Delta_o$ is detuned close to 2 GHz below the transition. This also allows the off-resonant probe beam to be completely transmitted through the medium. A single pass probe gain as high as 12 dB has been observed by heating the cell to a higher temperature. Experimentally, significant broadening of the gain linewidth has been observed either by increasing the pump laser intensity or by bringing the average frequency detuning $\Delta_o$ closer to the $5S_{1/2}$, F = 3 to $5P_{3/2}$, F′ = 4 $^{85}Rb$ resonance. A Raman gain doublet has been observed using bichromatic pump beams that are frequency separated by Δ = 2 MHz. These pump beams are generated using two AOMs driven by independent rf sources with frequencies 40 MHz and 42 MHz, respectively. Fig. 3 shows a time-averaged profile of the gain doublet observed as a function of the probe frequency, by varying the difference frequency δ $\equiv(\omega_2 - \omega_1)$ from the two photon resonance condition. Here, $\omega_2$ is defined as the probe frequency and $\omega_1$ as the average of the pump frequencies. During observation, the pump intensity is set equal to ~ 0.2 W/cm$^2$, which corresponds to a pump Rabi frequency of Ω ~ 4 GHz. The ratio of pump to probe intensities is set to about 20. A maximum gain of 3.5 dB has been observed in the background of the transmitted probe which varies due to the non-uniformity of the AOM diffraction across the probe scan. The frequency linewidth associated with the gain profiles is found to be ~ 700 KHz and the separation between the gain peaks is equal to δ. Satellite gain peaks around the main peaks are also observed due to additional frequency harmonics generated by our AOM drivers.

Fig. 4a shows Raman gain obtained with no time-averaging and under similar experimental conditions as above. The profile shows a large temporal modulation superimposed on the Raman gain spectrum, predominantly at the same frequency as the frequency separation between the pumps. Modulation at higher harmonic frequencies of Δ is also evident from the corresponding power spectrum, which are shown in fig. 5a. Fig. 4b shows a similar gain spectrum with an increased modulation frequency by using pump beams with Δ = 4 MHz. The magnitude of modulation was diminished with increasing pump separation due to the finite response time of the detector. In order to verify that the gain modulation is not due to residual pump leakage resulting from any kind of polarization rotation under the gain condition, we performed spectral measurements using a scanning Fabry-Perot (FP) resonator with an FSR ~ 17 GHz. The result of the measurement is shown in fig.6. The probe frequency is adjusted to the resonance condition for observing maximum gain. The dashed line in fig.6 shows a FP scan where we deliberately misaligned the polarization filtering to show the pump leakage for comparison. The solid line in fig.6 shows a scan with virtually no trace of the pump on the frequency-axis, corresponding to a position down-shifted by 3.0357 GHz with respect to the probe peak. The FP filter used for this purpose has a transmission of nearly 70%, thus confirming negligible pump leakage.

The presence of this temporal modulation may be interpreted as follows. The application of bichromatic pumps to a coherently driven two-level system produces a cascade of virtual excited levels with frequency separations that are integer multiples of the pump difference frequency $\Delta$. These virtual levels have different populations. The temporal modulation of the gain spectrum observed in our experiment results from the beating of the amplified probe beam with all coherently scattered Raman signals mediated through these virtual levels. A similar effect has also been discussed in the context of Bloch-Siegert oscillation and qubit rotation [16, 17]. Observations leading to similar modulation effects have also been reported in earlier experiments [18, 19] done using an atomic vapor medium.

The temporal modulation of the gain spectrum also causes modulation of the dispersion induced index profile, which follows from the Kramers-Kronig relation. This has been observed during dispersion measurements in our experiment. In general, index modulation may have a detrimental effect on the performance of a CAD based resonator gyroscope. However, it is actually possible to use this modulation to our advantage. In particular, it can be used to implement a dither modulation in order to overcome the so-called lock-in phenomenon [20] that occurs in a passive resonator from mode-coupling due to backscattering. A precise control over modulation amplitude would be required for this application. On the other hand, the modulation can be eliminated using an alternative experimental arrangement where the gain is produced by cascading two gain media (or vapor cells), each one being driven by a separate monochromatic pump field. In such a configuration, the probe still sees two gain peaks, and the concomitant dispersion. However, since the two pump frequencies do not interact with the same atoms, there is no temporal modulation of the gain profile.

As stated earlier, a key objective here is to measure accurately the magnitude of the linear dispersion between the two gain regions, while controlling it to achieve a value of $n_g$ close to zero. In order to measure the phase due to dispersion or the index variation resulting from Raman gain, we have used a heterodyne method (fig. 2) in our experimental setup. A non-resonant auxiliary or reference wave is produced by frequency shifting a fraction of the probe beam using a 40 MHz AOM. This is then divided in two parts: one is combined with the probe that experiences Raman gain and the other with the unperturbed fraction of the probe that does not propagate through the cell. These two heterodyne RF signals are detected using two fast photodetectors (response time < 10 ns). The phase difference between the two rf signals varies due to dispersion as the probe frequency scans around the gain resonance, and is given by $\delta\phi = k\,\Delta n(\omega)L$, where k is the magnitude of the wave vector and L is the length of the vapor cell. A low phase noise RF mixer and a low-pass frequency filter (dc to 300 KHz) are used to demodulate the rf signal from the detectors. The amplitude of the demodulated signal is proportional to $\delta\phi$ for $|k\,\Delta n(\omega)\,L| << 1$, which is valid if $|\Delta n| < 10^6$ (typical of dilute atomic medium). Unlike a homodyne approach, this technique is immune to phase drifts caused by external vibrations and allows us to measure accurately the dispersion in the atomic medium under gain resonance [21].

Fig.7 shows the dispersion associated with double Raman gain as calculated from the experimentally measured variation of the phase as a function of the difference frequency detuning $\delta$. A maximum index modulation $\Delta n$, which is of the order of $10^{-6}$ ,has been observed, with a steep negative dispersion slope ($\partial n/\partial \omega \sim 10^{-12}$ rad$^{-1}$sec, $n_g \sim -1000$). Large gain and dispersion slopes are easily achievable experimentally by increasing the coupling laser intensity and/or the cell temperature. Under experimental conditions, the index measurement is highly repeatable and is accurate to about 80%. Several factors such as electronic phase and frequency jitter, gain modulation and intensity noise limit the measurement accuracy. Fig. 8 shows an experimental measurement (non-averaged) where the dispersion slope is varied by changing the frequency separation between the pumps. The slope changes by more than two orders of magnitude before it is limited by the measurement accuracy. For a pump frequency separation $\Delta$ = 4 MHz in fig.8, a linear dispersion slope ~ 4 x $10^{-15}$ rad$^{-1}$ sec was measured over a bandwidth 0.5 MHz. This corresponds to a negative group index $n_g$ value of -8.66. Although our current measurement ability prohibits us from measuring extremely shallow dispersion slopes (4.1 x $10^{-16}$ rad$^{-1}$ sec) near the CAD condition, it clearly demonstrates the controllability to attain the dispersion condition desirable for the gyroscope operation. In order to determine the frequency separation $\Delta$ at which the $n_g$ can be close to zero, we have done a numerical extrapolation to the experimental data for linear dispersion slopes measured from fig. 8. Under the particular experimental conditions chosen for bi-frequency Raman gain, fig. 9 shows the variation in $n_g$ extrapolated as a function of Δ. Within our measurement accuracy, it shows that a Δ value close to 4.1 MHz can give rise to a value of $n_g$ that is nearly equal to zero. As mentioned earlier, the temporal modulation of the phase or index profile due to gain modulation from bi-frequency pumps has also been observed by removing the low-pass filter while demodulating the signal from the detectors. As expected, the bandwidth of the linear negative dispersion region decreases with increasing pump separation. The resulting limitation on the bandwidth of the operation of a CAD-enhanced gyro is not expected to be significant for practical applications.

In conclusion, we have investigated the resonant anomalous dispersion associated with bi-frequency Raman gain in Rb vapor for CAD enhanced rotation sensing. The dispersion slope is continuously tuned by varying the pump frequency separation, in order to achieve a group index close to the null value where the sensitivity enhancement is expected. The experiment also shows that the linearity and the dynamic range of dispersion which govern the performance of a CAD-enhanced gyroscope can be controlled by easily accessible experimental parameters. As such, the work presented here represents a key step in establishing the feasibility of a CAD-enhanced gyroscope.

This work was supported in part by the AFOSR and the ARO MURI program.

**List of Figures:**

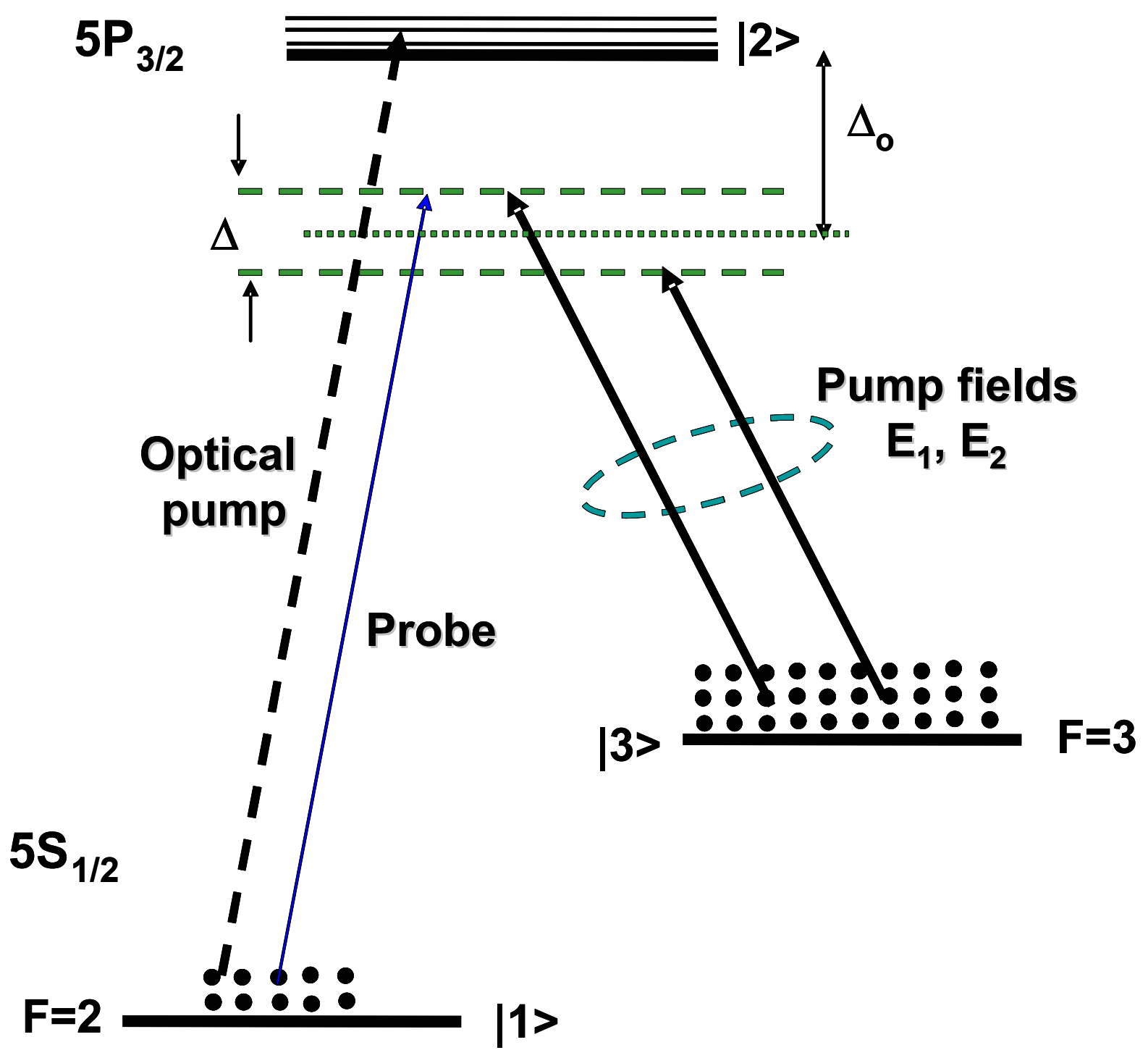

**Figure 1.** Energy level diagram showing optical excitations in [85]Rb D2 transitions for stimulated bi-frequency Raman gain

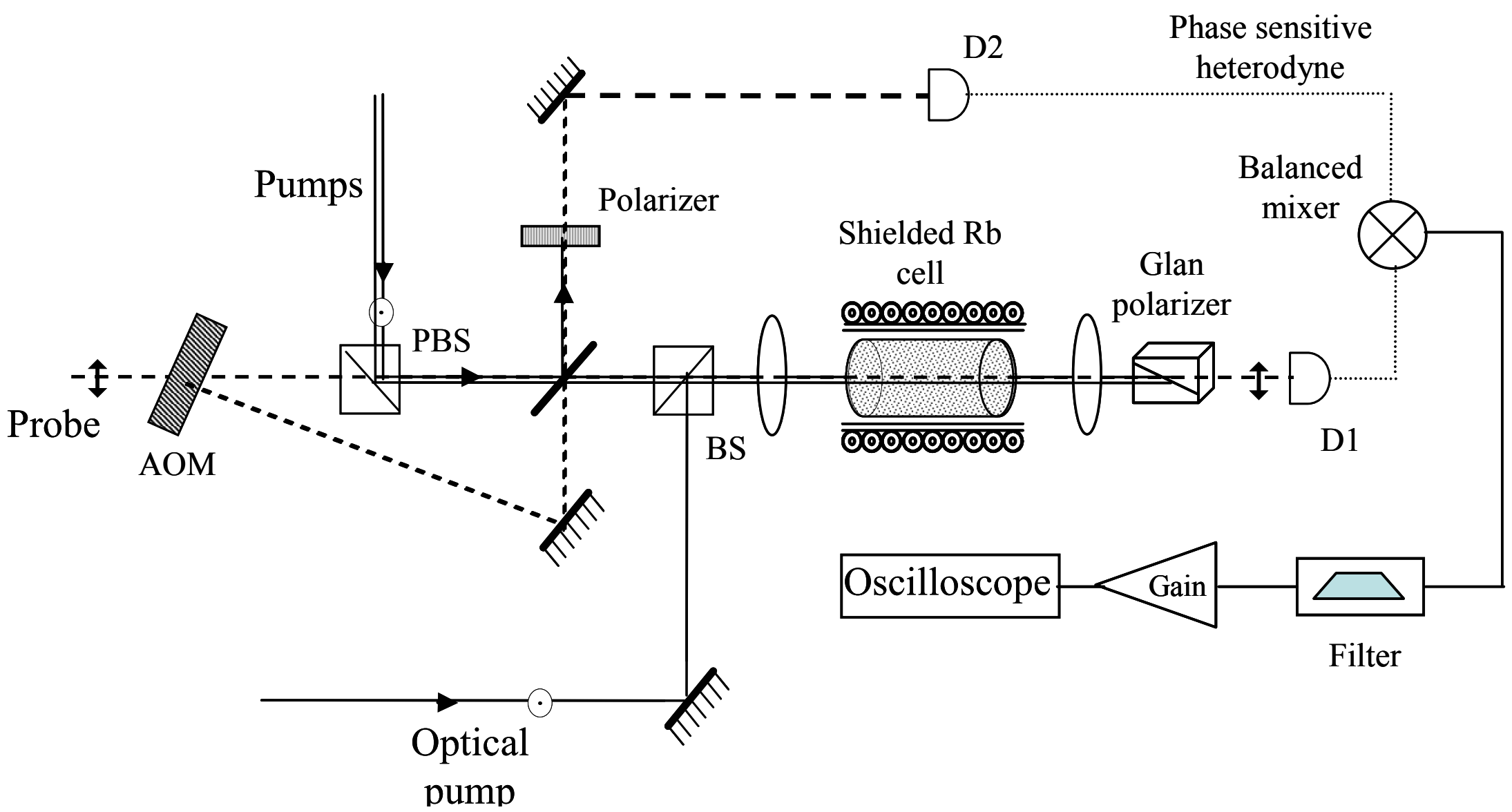

**Figure 2** Experimental arrangement. PBS, polarizing beam splitter, AOM, acousto-optic modulator, D, photodiode

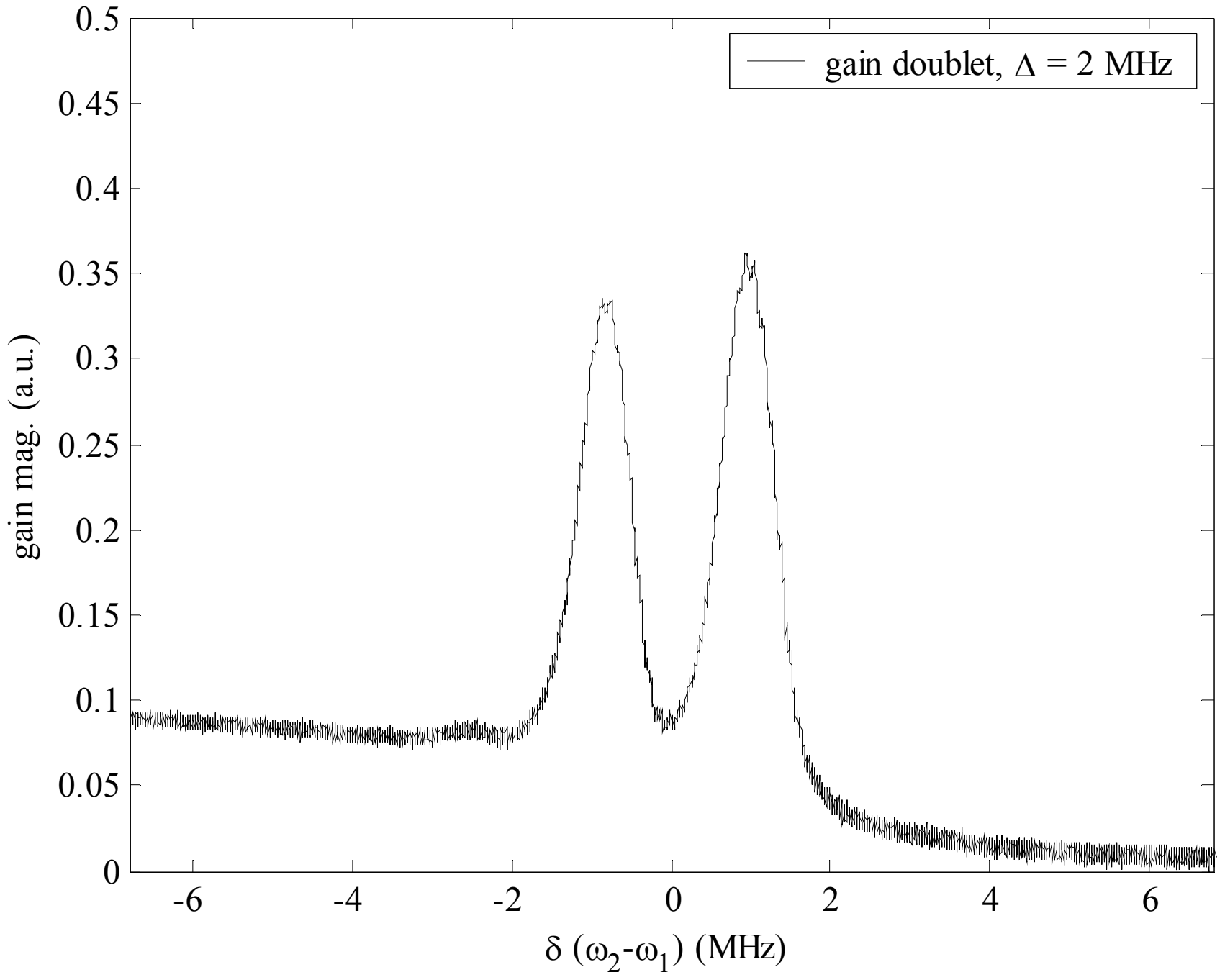

**Figure 3** Time-averaged Raman gain doublet using bi-frequency pumps with $\Delta = 2$ MHz.

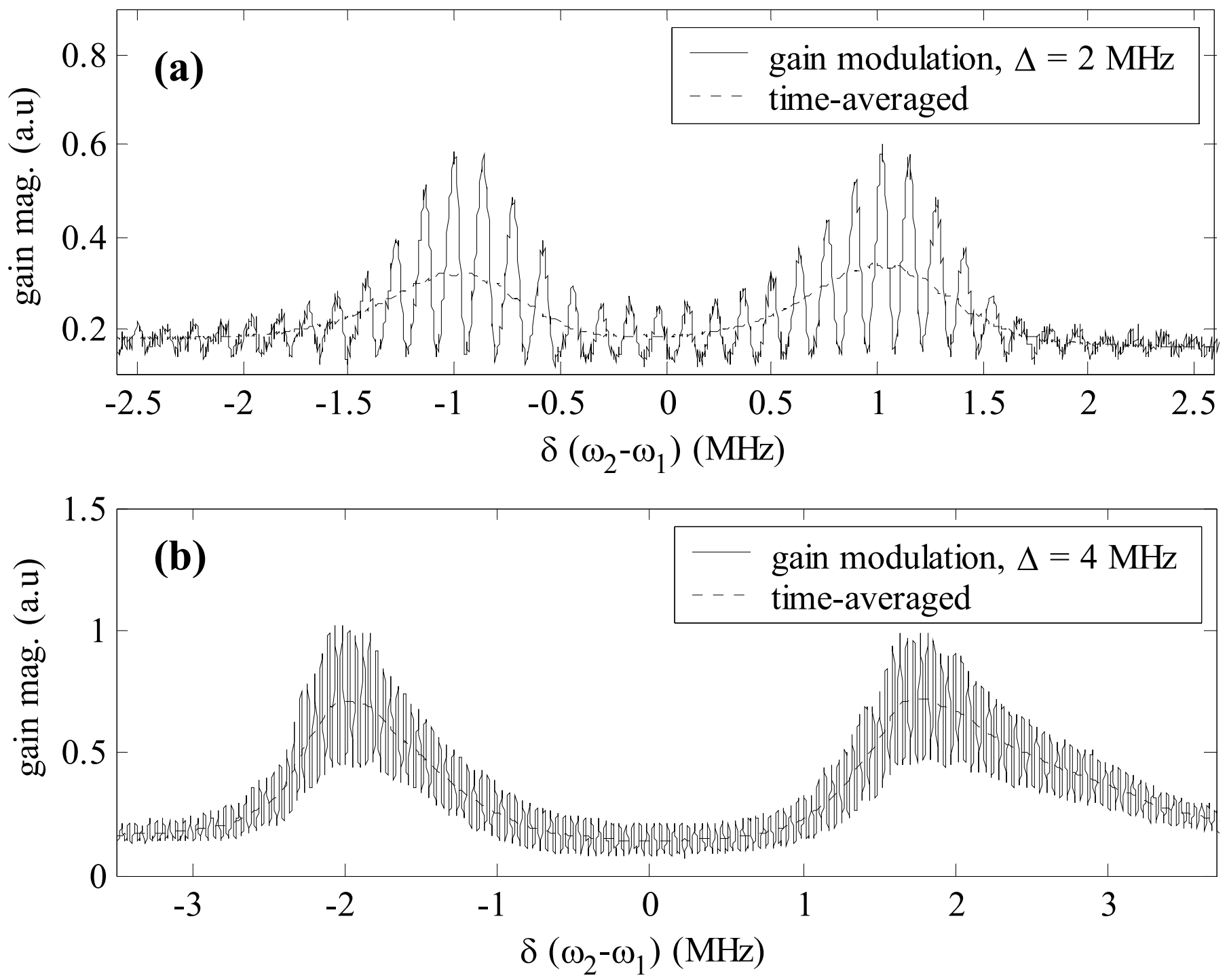

**Figure 4** Gain modulation using bi-frequency pumps with (a) $\Delta = 2$ MHz

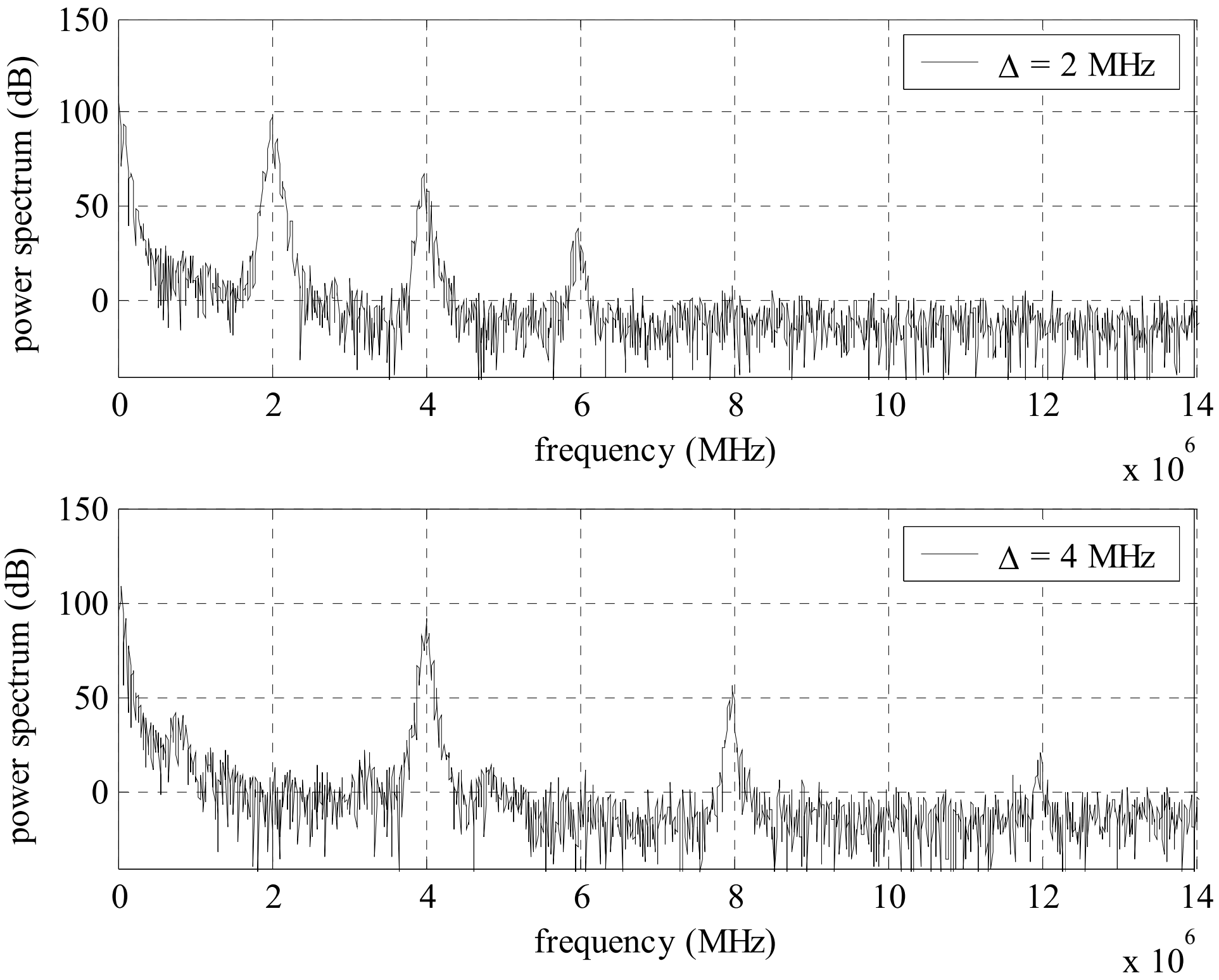

**Figure 5** Power spectrum of gain doublet showing modulation frequencies at multiples of pump frequency separation for $\Delta$ = (a) 2 MHz (b) 4 MHz.

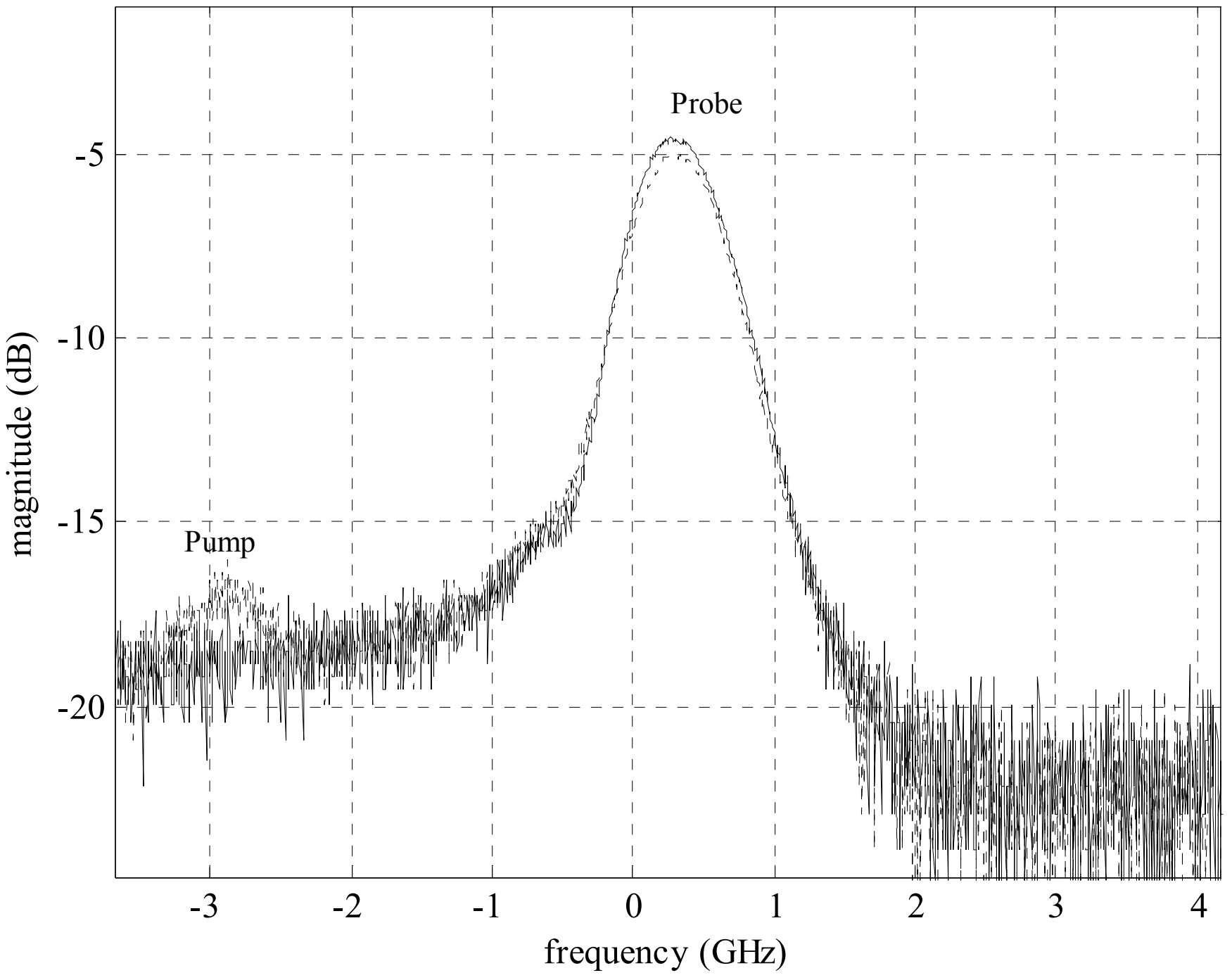

**Figure 6** Spectrum showing transmitted probe and pump through a scanning FP filter at maximum probe gain. Negligible pump leakage (solid line) is observed compared to the case when polarization filter is

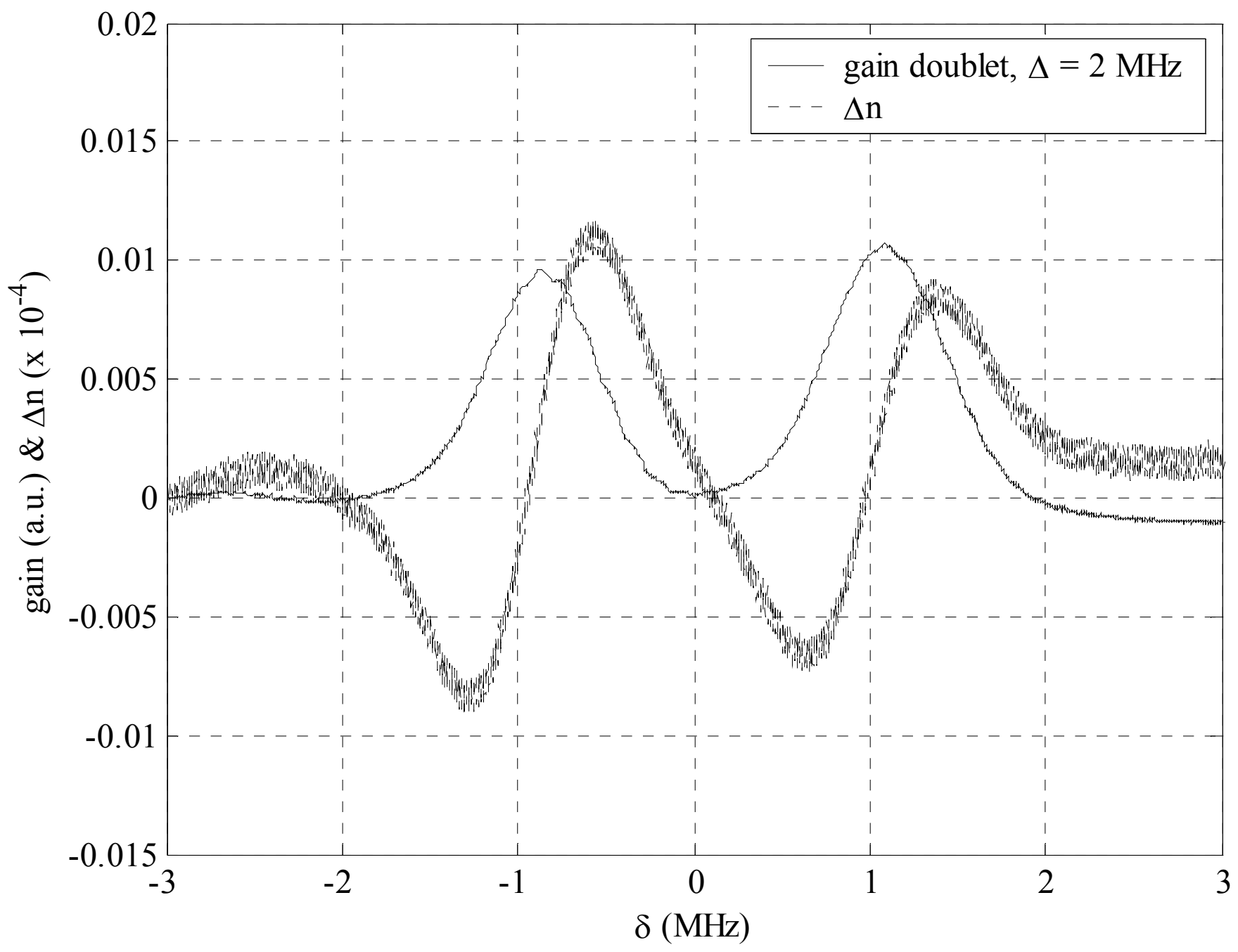

**Figure 7** Measured dispersion ($\partial n/\partial\omega \sim -2.65 \times 10^{-13}$ rad sec$^{-1}$, $n_g \sim 639$) associated with bi-frequency Raman gain using the heterodyne technique.

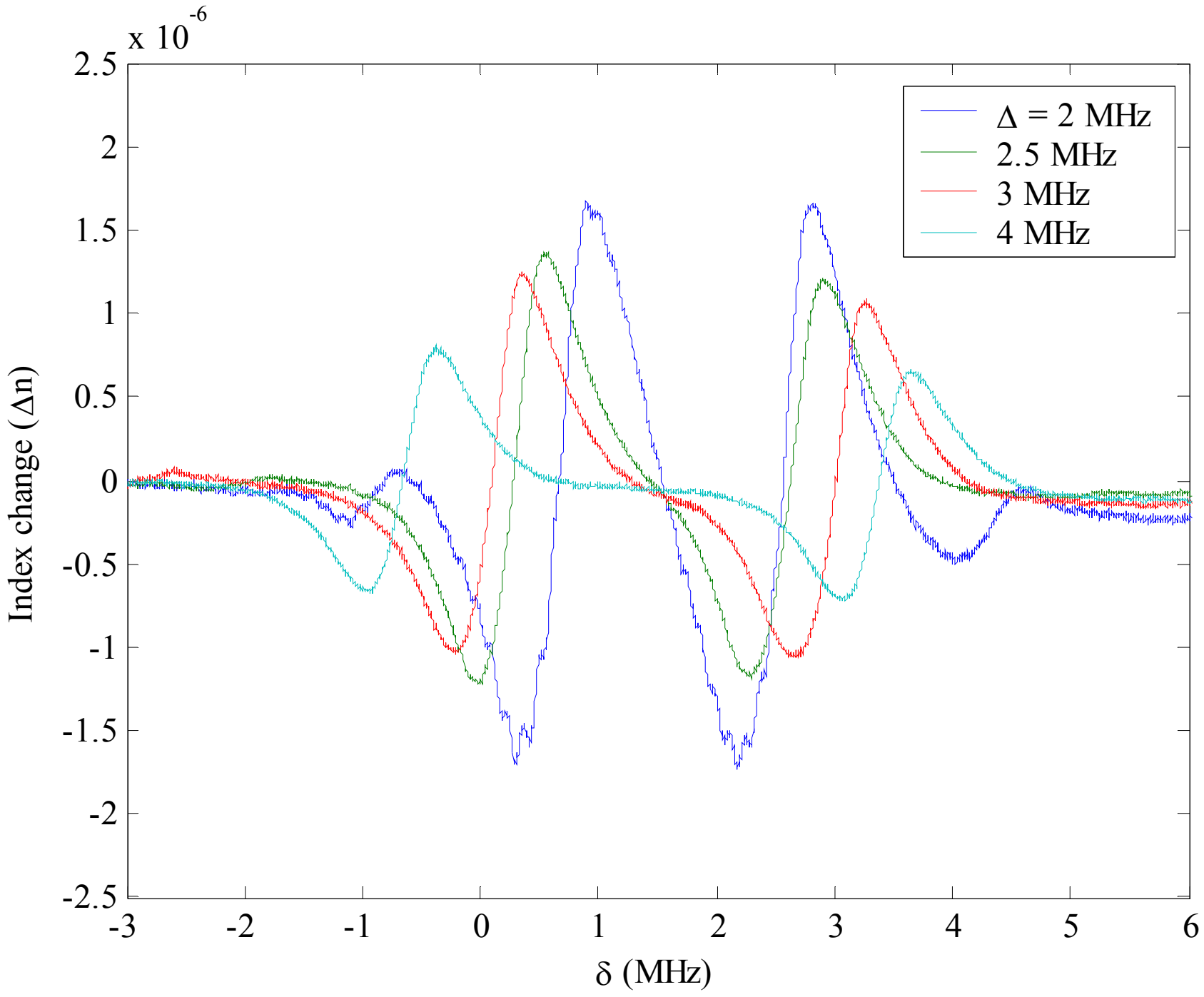

**Figure 8** Heterodyne measurement mapping variation in dispersion slope with pump separation Δ (a) 2 MHz, $\partial n/\partial\omega = -1.08\mathrm{x}10^{-12}$ $rad^{-1}$ sec, $n_g = -2608$ (b) 2.5 MHz, $\partial n/\partial\omega = -1.4\mathrm{x}10^{-13}$ $rad^{-1}$ sec, $n_g = -337.3$ (c) 3 MHz, $\partial n/\partial\omega = -8.05\mathrm{x}10^{-14}$ $rad^{-1}$ sec, $n_g = -193.5$ (d) 4 MHz, $\partial n/\partial\omega = -4\mathrm{x}10^{-15}$ $rad^{-1}$ sec, $n_g = -8.66$. These measurements were done over a bandwidth $\Delta f = 0.5$ MHz.

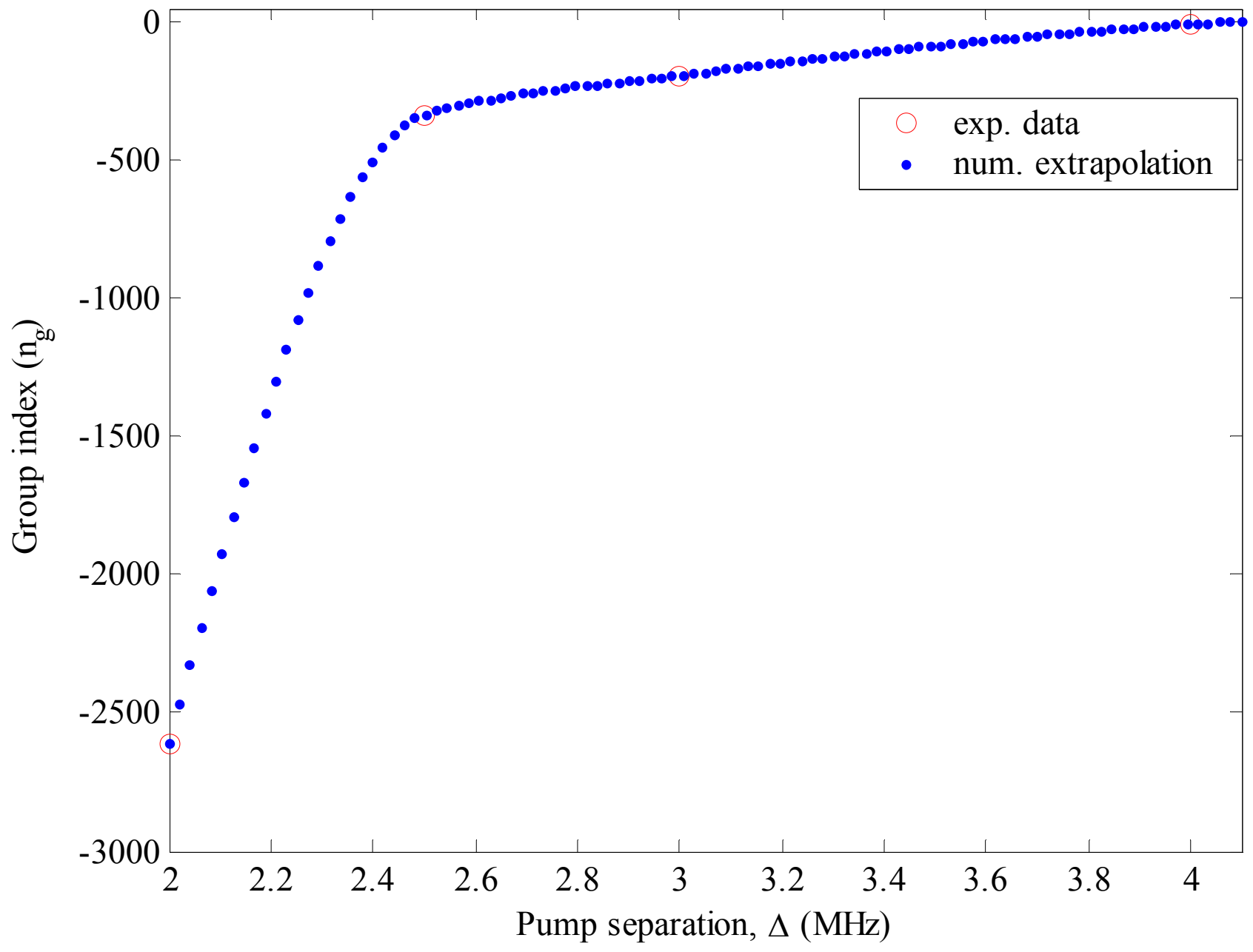

**Figure 9** Group index ($n_g$) change with pump frequency separation (Δ) estimated from dispersion profiles shown in fig. 8. A numerical extrapolation of experimental data shows that group index null ($n_g$ =0) can be achieved using Δ~4.1 MHz, for the parameters chosen under experimental condition.